\pdfoutput=1

\documentclass[a4paper,11pt]{article}
\usepackage{jcappub} 
\usepackage{color}
\usepackage{xcolor}
\usepackage{amsmath}
\usepackage{amsfonts}
\usepackage{graphicx} 
\usepackage{amssymb}
\usepackage{epstopdf}
\usepackage{url}
\usepackage{bm}
\usepackage{hhline}
\usepackage{multirow}
\usepackage{epsfig}
\usepackage{varioref}
\usepackage{amssymb}
\usepackage{comment} 
\usepackage{cases}
\usepackage{multirow}
\usepackage{mathrsfs}
\usepackage{calrsfs}
\usepackage{mathtools}
\usepackage{amsthm}
\usepackage{hyperref}
\usepackage{wrapfig}
\usepackage{caption}
\usepackage{subcaption}

\newcommand{\PR}[1]{\ensuremath{\left[#1\right]}} 
\newcommand{\PC}[1]{\ensuremath{\left(#1\right)}} 

\include{configs} 

\title{On the phenomenology of extended Brans-Dicke gravity}
\author{Nelson A. Lima $^{1}$} 
\author{and Pedro G. Ferreira $^{2}$}
\affiliation{$^{1}$Institute for Astronomy, University of Edinburgh, Royal Observatory, Edinburgh EH9~3HJ, United Kingdom}
\affiliation{$^{2}$Astrophysics, University of Oxford, Denys Wilkinson Building, Keble Road, Oxford, OX1 3RH, UK}

\emailAdd{ndal@roe.ac.uk, p.ferreira1@physics.ox.ac.uk}
\date{\today}

\abstract{
We introduce a designer approach for extended Brans-Dicke gravity that allows us to obtain the evolution of the scalar field by fixing the Hubble parameter to that of a $w$CDM model. We obtain analytical approximations for $\phi$ as a function of the scale factor and use these to build expressions for the effective Newton's constant at the background and at the linear level and the slip between the perturbed Newtonian potentials. By doing so, we are able to explore their dependence on the fundamental parameters of the theory. }

\begin{document}
\maketitle

\section{\label{Int}Introduction}

Over the next decade, we expect a step change in our understanding of gravity on cosmological scales. Surveys of large scale structure should be able to pin down the expansion of the Universe and the growth of structure with exquisite precision \cite{euclid,lsst,wfirst,ska}. These new data sets should allow us to constrain modifications to general relativity at a level which may be comparable to those obtained on astrophysical scales.

If we are to fully take advantage of these data sets, it is essential to have a detailed and accurate understanding of how different observables depend on our assumptions about gravity. In particular, we should know how deviations from general relativity will affect our observations: whether the effects are large or small (given what we know on astrophysical scales) and how correlations between the observables themselves might be indicative of some underlying structure.

There has been a formidable campaign to develop methods for studying the effects of modified gravity on large scales (for a compendium of theories, see \cite{reviewall}). A different approach has been to develop a unified method of parameterizing all possible theories at the linearized level (for a selection of methods, see \cite{tessapar,battye,gubitosi,gleyzes}). Yet, while there is an inexorable momentum that should lead to a battery of effective techniques for extracting useful information from the data, we do not have yet a firm understanding of what to expect. By this we mean that, given certain theoretical assumptions, what our observables should look like, i.e. what values should they take and how should they be interrelated as a function of whatever fundamental parameters we might consider.

In principle, the step from taking the parameters, $\alpha_i$ (with $i=1,\cdots N$), of some underlying theory and working out the resulting phenomenological parameters, $\beta_j$ (with $j=1,\cdots M$) tied to observations, should be straightforward. In practice, the process can be complicated, highly non-linear, degenerate and normally obscures the relationship between the prior assumptions on $\alpha_i$ and the resulting theoretical priors on  $\beta_j$. One way around this is to develop an approximate mapping between the two sets of parameters and, wherever possible, analytic relations between the two. Furthermore, if one can find a method for restricting the range of $\alpha_i$ given some assumptions about a subset of the $\beta_i$, one can quickly surmise what correlations and covariance one should expect for the remaining phenomenological parameters. In this paper we propose an approach to do so, considering a restricted  model for cosmological modifications to gravity. 

Our starting point is a well known theory, the Brans-Dicke (BD) theory of gravitation \cite{bd1}. This theory is the simplest scalar-tensor theory one can envisage \cite{st1,st2,st3,st4,st5} and is considered a viable alternative to General Relativity, one which respects Mach's Principle. Since its formulation, this theory has been exhaustively studied as a possible alternative solution for the accelerated expansion of the Universe.

It has been shown that Brans-Dicke theory can produce accelerated solutions for small, negative values of the BD parameter $\omega_{\rm{BD}}$ \cite{bd2,bd6}. Given that one recovers standard GR in the limit where $\omega_{\rm{BD}} \rightarrow \infty$, such values of the $\omega_{\rm{BD}}$ clash with Solar system constraints \cite{will,bertotti}; furthermore, recent constraints with the latest CMB data are also not compatible with such low values of $\omega_{\rm{BD}}$ \cite{planckbd,planckbd2}. Several modifications of this theory try to include self-interacting potentials \cite{bd3,bd4,bd7} or consider a field-dependent Brans-Dicke parameter $\omega(\phi)$ \cite{bd5}, without solving this problem successfully. Also, models with a non-minimal coupling of the scalar field have been considered in Refs.~\cite{bd8,bd9,bd10,bd11}.

In this paper we construct a theory of {\it designer}, extended Brans-Dicke gravity and use it to characterize the form of the observables we might measure. This theory is "extended" because we include a potential for the Brans-Dicke field and we dub it "designer" (the term ``designer'' was first used in models of inflation that attempted to match observations by designing the density fluctuation spectra \cite{infdesigner}) because we reconstruct the potential (which might not have an analytic form) from a desired background evolution. While such a theory does not seem fundamental, it might be seen as an approximation to a scalar-tensor theory which has a particular, a priori, form of the background evolution. Our construction allows us to find a number of analytic approximations and, in doing so, lets us gain a firmer understanding of the phenomena we want to study. 

Our designer approach for the extended Brans-Dicke gravity is novel. It allows us to retrieve the evolution of the scalar field, $\phi$, by fixing the background evolution and is robust for high values of the BD parameter, which is the regime we are interested in. This method not only works for a $\Lambda$CDM like evolution with an effective equation of state $w_{\rm{eff}} = -1$, but is also applicable for models with $w_{\rm{eff}} > -1$ as in a $w$CDM scenario. And, for both cases, we are able to retrieve analytical approximations for $\phi$ as a function of the scale factor $a$ which could prove useful for a faster and more efficient fitting of models to data. 

The paper is structured as follows. In Sec.~\ref{intro} we introduce the Brans-Dicke theory with a constant $\omega_{\rm{BD}}$ parameter. In Sec.~\ref{designerbd} we describe the designer approach, motivated by an analysis of the behavior of this theory when we have a constant potential $V(\phi)$. In Sec.~\ref{analytical} find approximate analytic solutions to the evolution of the scalar field and use it to infer the shape of the potential. We then use these results in Sec.~\ref{phenom} to construct analytical approximations to the phenomenological parameter which can be constrained by data. In Sec.~\ref{discussion} we discuss our results.

\section{\label{intro}Extended Brans-Dicke gravity: background equations}

The action for extended Brans-Dicke theory in the Jordan frame, is given by
\begin{equation}{\label{bdaction}}
 S = \frac{1}{2\kappa^{2}}\int d^{4} x \sqrt{-g} \PC{\phi R - \frac{\omega_{\rm{BD}}}{\phi}\PC{\partial \phi}^{2} - 2 V(\phi)} + S_{\rm{m}},
\end{equation}
\noindent where $S_{\rm{m}}\PR{\Psi_{\rm{m}};g_{\mu \nu}}$ is the minimally coupled matter Lagrangian and $\kappa^2 = 8 \pi G$, where $G$ is Newton's gravitational constant measured today. Varying the action with respect to the metric elements, we find the Einstein equations,
\begin{equation}{\label{einstein}}
 G_{\mu \nu} = \frac{\kappa^2}{\phi} T_{\mu \nu}^{\rm{m}} + \frac{\omega_{\rm{BD}}}{\phi^2}\PR{\phi_{,\mu}\phi_{,\nu} - \frac{1}{2}g_{\mu\nu}\phi_{,\alpha}\phi^{,\alpha}} + \frac{1}{\phi}\PR{\phi_{,\mu;\nu}-g_{\mu \nu} \Box \phi} - \frac{V(\phi)}{\phi} g_{\mu \nu},
\end{equation}
\noindent where $T_{\mu \nu}^{\rm{m}}$ is the matter stress-energy tensor.

By varying the action (\ref{bdaction}) with respect to the field, one gets the field's equation of motion
\begin{equation}{\label{fieldmotion}}
 \Box \phi = \frac{ \kappa^2 T}{3 + 2 \omega_{\rm{BD}}} - \frac{2}{3 + 2 \omega_{\rm{BD}}}\PR{2 V(\phi) - \phi V_{\phi}},
\end{equation}
\noindent where $V_{\phi} \equiv dV/d\phi$. Considering a flat Friedmann-Lemaitre-Robertson-Walker (FLRW) metric, $ds^2 = -dt^2 + a^{2}(t)d\vec{x}^2$, this equation reads
\begin{equation}{\label{fieldfrw}}
 \ddot{\phi} + 3 H \dot{\phi} = \frac{\kappa^2 \rho_{\rm{m}}}{3+2 \omega_{\rm{BD}}} + \frac{4V(\phi) - 2\phi V_{\phi}}{3+2 \omega_{\rm{BD}}},
\end{equation}
\noindent where $\rho_{\rm{m}}$ is the matter's energy density and $H \equiv \dot{a}/a$ is the Hubble parameter. The latter is determined by the two Friedmann equations, which are written as
\begin{eqnarray}{\label{friedmann}}
3 H^{2} \phi &=& \kappa^2 \rho_{\rm{m}} - 3H \dot{\phi} + \frac{\omega_{\rm{BD}}}{2}\frac{\dot{\phi}^{2}}{\phi} + V(\phi) \\
2\dot{H} + 3H^{2} &=& -\kappa^2\frac{p_{\rm{m}}}{\phi} - \frac{\omega_{\rm{BD}}}{2}\frac{\dot{\phi}^{2}}{\phi^2} - 2H \frac{\dot{\phi}}{\phi} - \frac{\ddot{\phi}}{\phi} + \frac{V(\phi)}{\phi}. \nonumber
\end{eqnarray}

Lastly, from the previous equations, one can define an effective equation of state for the dark energy component of our model, which is given by
\begin{equation}{\label{weff}}
 w_{\rm{eff}} = \frac{\dot{\phi}^{2} \omega(\phi) + 4 H \dot{\phi} + 2 \ddot{\phi} - 2 V(\phi)}{\dot{\phi}^{2} \omega(\phi) - 6 H \dot{\phi} + 2 V(\phi)},
\end{equation}
\noindent where $\omega(\phi) = \omega_{\rm{BD}}/\phi$ and, even more straightforwardly, one can define the fractional effective dark energy density parameter,
\begin{equation}{\label{fraceff}}
 \Omega_{\phi} = \frac{\rho_{\phi}}{3 H^2 \phi},
\end{equation}
\noindent where the effective energy density is given by
\begin{equation}{\label{rhoeff}}
 \rho_{\phi} = \frac{\omega_{\rm{BD}}}{\phi}\frac{\dot{\phi}^{2}}{2} - 3 H \dot{\phi} + V(\phi).
\end{equation}

\subsection{\label{constantpot}Constant Potential V($\phi$)}

Before proceeding to the designer approach, we can get an idea of the different effects at play in extended Brans-Dicke gravity by considering the case of a constant potential $V(\phi)$. For all our calculations in this section, we have $V(\phi) = 3 H_{0}^{2} \PC{1 - \Omega_{\rm{m}}} \equiv V$, where $\Omega_{\rm{m}}$ is the fractional present-day energy density of matter. For a perfect $\Lambda$CDM scenario we should have an effective dark energy equation of state equal to $-1$ during the whole cosmological evolution, with the scalar field remaining perfectly still and showing no evolution at all. However, in the Brans-Dicke paradigm, the field should always evolve even if its dynamics are subdominant (in ``slow roll") compared to the potential $V$. Hence, effectively, we will have a quasi-$\Lambda$CDM evolution.

We start by numerically solving the scalar field evolution using Eqs.~(\ref{fieldfrw}) and (\ref{friedmann}) considering a constant potential as defined in the previous paragraph. We set the initial conditions for the scalar field deep within the matter dominated regime at a redshift around $z_{\rm{i}} \approx 1000$. For this, we consider a known solution of Brans-Dicke gravity given by \cite{pl1,pl2,pl3}
\begin{equation}{\label{attphi}}
 \phi = \phi_{0} a^{1/\PC{\omega_{\rm{BD}}+1}},
\end{equation}
\noindent where $\phi_{0} = \PC{2\omega_{\rm{BD}}+4}/\PC{2\omega_{\rm{BD}}+3}$. This solution is, in fact, an attractor solution of the system derived in the absence of a potential $V(\phi)$ and for a Universe dominated by matter alone \cite{pl1,pl2,pl3}. The scale factor, on the other hand, evolves as \cite{pl1,pl2,pl3}
\begin{equation}{\label{atta}}
 a(t) = \PC{\frac{t}{t_0}}^{\PC{2\omega_{\rm{BD}}+2}/\PC{3\omega_{\rm{BD}}+4}},
\end{equation}
\noindent and we see that, in the GR limit of $\omega_{\rm{BD}} \rightarrow \infty$, $\phi = 1$ and $a(t) \propto t^{2/3}$ throughout the matter dominated regime; $t_{0}$ is related to the inverse of the present-day value of the Hubble parameter, $H_{0}$, such that $t_{0} H_{0} = \PC{2 \omega_{\rm{BD}}+2}/\PC{3 \omega_{\rm{BD}}+4}$. The value of $\phi_{0}$ ensures that, in a matter dominated Universe, we would measure an effective gravitational constant today, $G_{\rm{eff}}$, equal to the actual Newton's gravitational constant, $G$, in Cavendish-like experiments. This assumes, of course, that the Solar system value of $\phi$ is representative of the Universe as a whole, which may not be entirely accurate \cite{cliftonsolar}.

Let us also point out that, in a matter dominated flat Universe, the matter density will not be precisely equal to the critical density due to a very small, negative, and almost negligible contribution from the scalar field dynamics. It is possible to rescale the matter density (as in Ref.~\cite{andrew}), but we opt not to do so, since the correction is negligible in the  $\omega_{\rm{BD}} >> 1$ regime we are mostly interested in this work.

\begin{figure}[t!]
\begin{center}
\includegraphics[scale = 0.50]{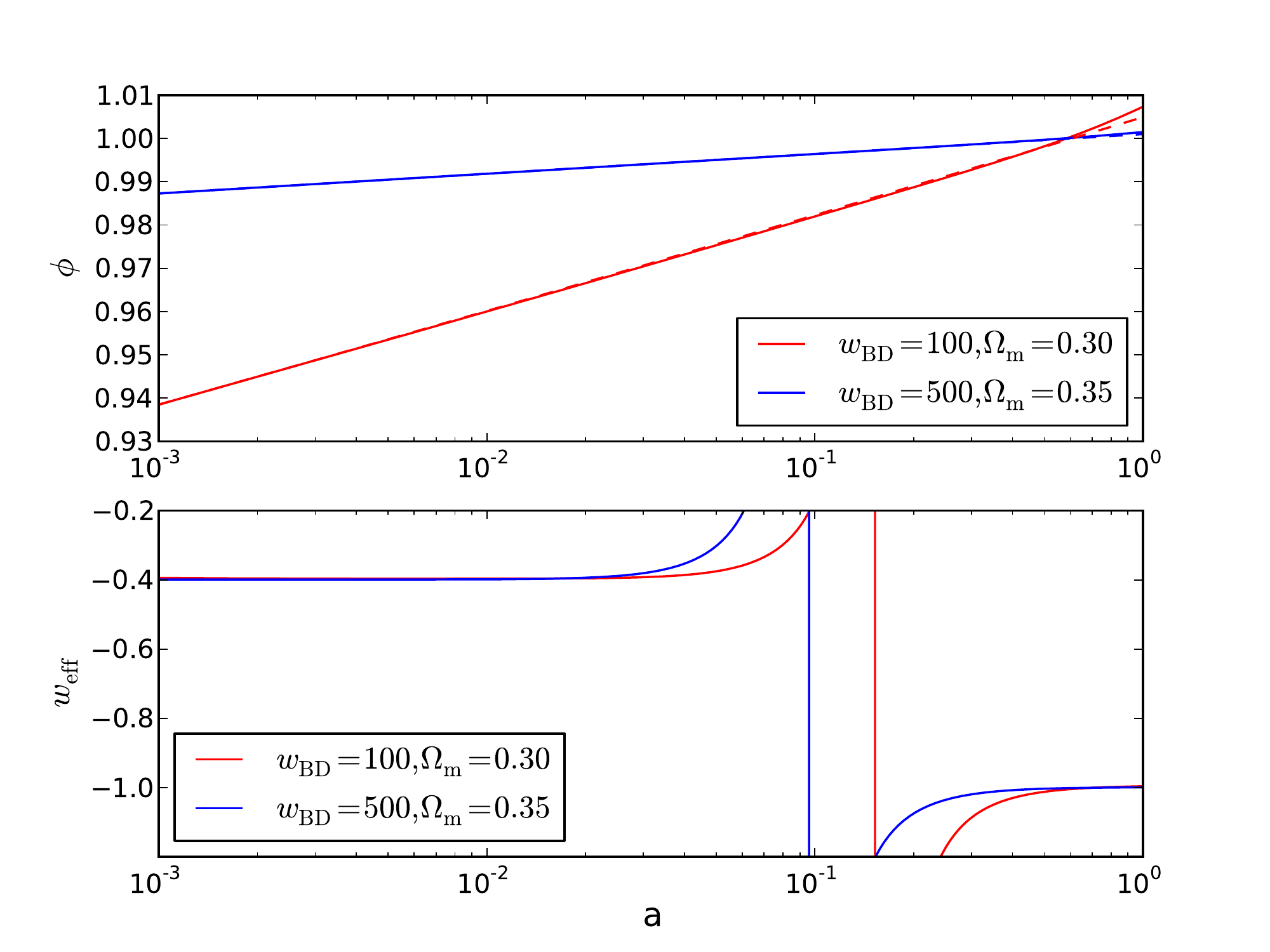}
\caption{\label{figure1} We plot the numerical evolution (solid lines) of the scalar field evolution and the effective dark energy equation of state, $w_{\rm{eff}}$ in the presence of a constant potential. We note that, in the matter dominated regime, $\phi$ evolves according to a known power-law solution given by Eq.~(\ref{attphi}), which we also plot (dashed lines).}
\end{center}
\end{figure}

In Fig.~\ref{figure1}, we have the numerical evolution of the scalar field plotted against the power-law solution given by Eq.~(\ref{attphi}). We can clearly observe that, even in the presence of a constant potential $V$, the Brans-Dicke scalar field evolves according Eq.~(\ref{attphi}) at early-times, during the matter dominated epoch. Only at late-times, close to $a = 1$, we see a slight departure from the power-law of Eq.~(\ref{attphi}), when the dark energy component begins to dominate and accelerates the scalar field.

Still in Fig.~\ref{figure1} we can observe the numerical evolution of the dark energy effective equation of state $w_{\rm{eff}}$ as given by Eq.~(\ref{weff}). We observe a very sharp transition from $-0.4$ to $-1$ that we will explain later on. For now, we can conclude that, even though the scalar field is accelerated by the presence of the constant potential $V$, its dynamics remain subdominant (the aforementioned slow roll evolution) and allow for a late-time potential dominated epoch with $w_{\rm{eff}} = -1$

Having shown in Fig.~{\ref{figure1}} that we recover the power-law solution given Eq.~(\ref{attphi}) at early-times, we now extend its application by using it in the effective equation of state $w_{\rm{eff}}$ given by Eq.~(\ref{weff}) in the presence of a constant potential $V$. Hence, we approximately obtain
\begin{equation}{\label{weffatt}}
 w_{\rm{eff}} \approx \frac{4 - 4 \omega_{\rm{BD}} V a^{3}/H_{0}^{2}}{-10 + 4 \omega_{\rm{BD}}V a^{3}/H_{0}^{2}},
\end{equation}
\noindent in the limit of $\omega_{\rm{BD}} >> 1$, and where we have also used Eq.~(\ref{atta}). Hence, in the matter dominated regime, the potential contribution is suppressed by the scale factor leading to $w_{\rm{eff}} \approx -0.4$ (unless $\omega_{\rm{BD}} \rightarrow \infty$ and $V \ne 0$). Thus, for values of $\omega_{\rm{BD}}$ which are consistent with Solar System constraints, it is impossible to get an accelerated solution without adding a potential $V(\phi)$, that may not necessarily be constant. However, with a constant potential $V(\phi)$, one gets $w_{\rm{eff}} = -1$ at late times after a sharp, non-smooth transition from $w_{\rm{eff}} \approx -0.4$, which we have seen in Fig.~\ref{figure1}.

An effective equation of state $w_{\rm{eff}} \approx -0.4$ at early times could constitute a problem, eventually compromising the extension of the matter dominated regime and rendering the model inviable. However, calculating $\Omega_{\phi}$, given by Eq.~(\ref{fraceff}), explicitly during the matter dominated regime using Eqs.~(\ref{attphi}) and (\ref{atta}), one gets
\begin{equation}{\label{rhoeff_attractor}}
 \Omega_{\phi} \approx \frac{1}{3}\PR{-\frac{5}{2 \omega_{\rm{BD}}} + \frac{V(\phi)}{H_{0}^{2}}a^{3}},
\end{equation}
\noindent which, for large values of $\omega_{\rm{BD}}$ is negligible at early times. 

Also, we note that the discontinuity in $w_{\rm{eff}}$ happens due to a zero crossing of the denominator of Eq.~(\ref{weff}). If we change from physical time $t$ to the natural logarithm of the scale factor, $dt \rightarrow d\ln a$, we have that $d/dt \rightarrow H d/d\ln a$. Therefore, neglecting the $\phi^{\prime 2}$ (the prime denotes a derivative with respect to $\ln a$) term because this is proportional to $(1+\omega_{\rm{BD}})^{-2}$ before the transition and for large $\omega_{\rm{BD}}$, the denominator of $w_{\rm{eff}}$ can be approximated to just $-3\phi^{\prime} + V(\phi)/H^2$. Therefore, given that, in the matter dominated regime, $V(\phi)/H^2 \propto V(\phi) a^{3}/H_{0}^{2}$ is an increasing function of the scale factor, there will come a point at which this term will be equal to $3 \phi^{\prime}$, leading to the discontinuity in $w_{\rm{eff}}$. For the constant potential, the scale factor of the discontinuity is apprximately
\begin{equation}{\label{adisc}}
a_{\rm{disc}} \approx \PC{\frac{\Omega_{\rm{m}}} {1-\Omega_{\rm{m}}} \frac{1}{1+\omega_{\rm{BD}}}}^{1/3}. 
\end{equation}
The discontinuity in $w_{\rm{eff}}$ has no impact on the background expansion of the model: if we take the second Friedmann equation and $p_{\rm{m}} = 0$, we have
\begin{equation}{\label{secondfriedmann}}
 \frac{\ddot{a}}{a} = -\frac{H^{2}}{2}\PC{1 + 3\frac{w_{\rm{eff}}}{\rho_{\rm{m}}/\rho_{\phi}+1}}.
\end{equation}
Since the divergence in $w_{\rm{eff}}$ happens due to $\rho_{\phi}$ crossing zero, as we just discussed, no divergence is seen in the evolution of $\ddot{a}$ because the term $\rho_{\rm{m}}/\rho_{\phi}$ follows the behavior of $w_{\rm{eff}}$.

Finally, only when $\omega_{\rm{BD}} \rightarrow \infty$ (the General Relativity limit) does one get $w_{\rm{eff}} = -1$  throughout the whole evolution, as seen in Fig.~\ref{figure1}. Here the potential $V(\phi)$ will dominate and the scalar field dynamics is heavily suppressed. The discontinuity in $w_{\rm{eff}}$ will now happen at a much earlier time, as is clear from Eq.~(\ref{adisc}), leading to a smooth $w_{\rm{eff}} = -1$ in the case of a constant potential.

\section{\label{designerbd}Designer extended Brans-Dicke gravity}
Having presented the general form for extended Brans Dicke gravity, we now proceed to construct an algorithm that will lead to a particular expansion rate or, more specifically, to an effective equation of state. Hence, effectively, we design and impose the background history we wish for our model which in turn determines the dynamical evolution of the Brans-Dicke scalar field. We note that the authors of Ref.~\cite{acquaviva} suggested the designer method we will describe further, but did not fully explore its consequences.

\begin{figure*}
\begin{center}
$
\begin{array}{cc}
\includegraphics[scale = 0.385]{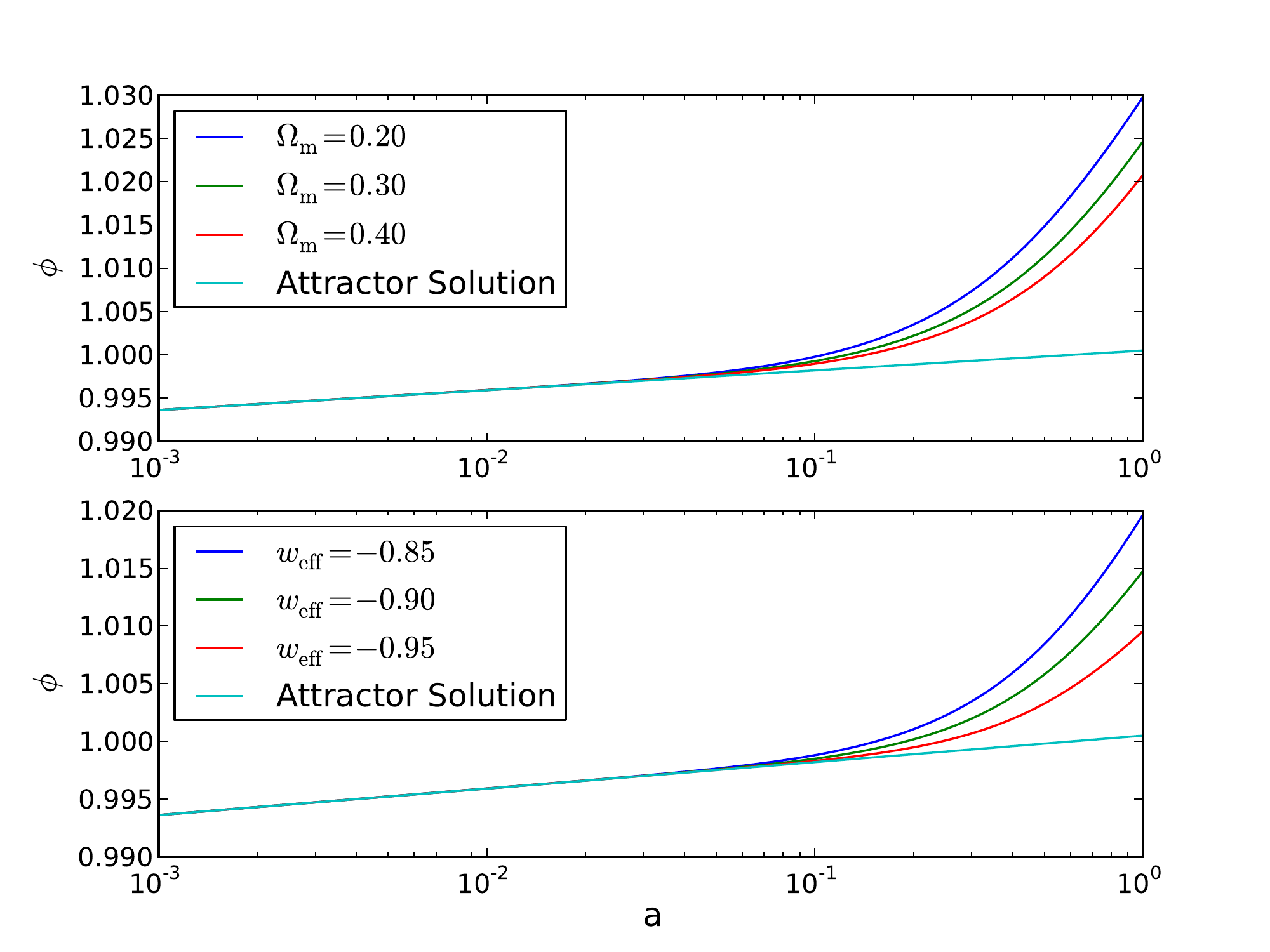} & \includegraphics[scale=0.385]{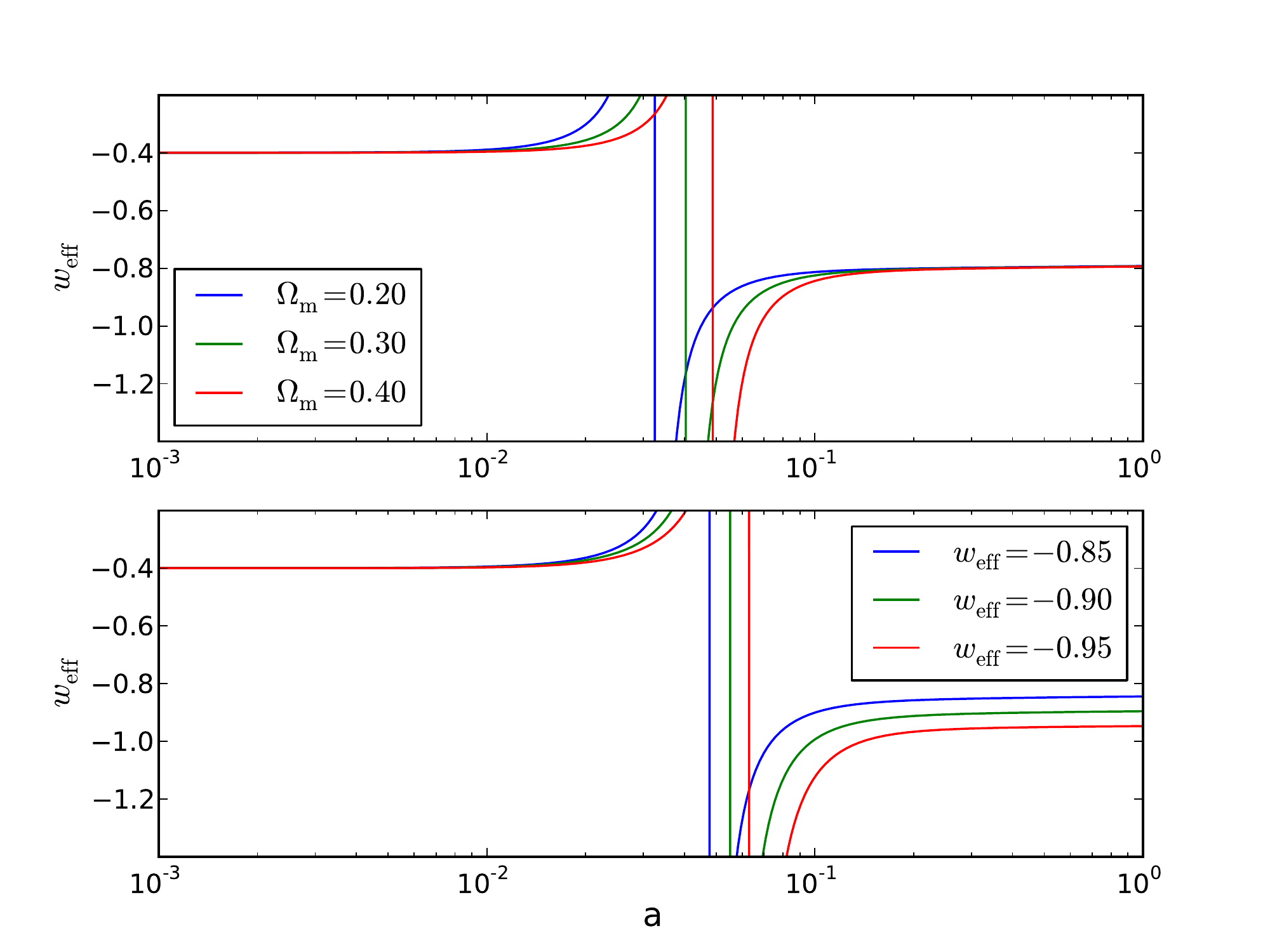}
\end{array}$
\caption{\label{figure2} We show the evolution of $\phi$ and $w_{\rm{eff}}$ as a function of the scale factor, for $\omega_{\rm{BD}} = 1000$. On the left, the top plot has $w_{\rm{eff}} = -0.80$ today, while the bottom plot has $\Omega_{\rm{m}} = 0.30$ and different values of $w_{\rm{eff}}$ today. On the right we show the respective evolution of $w_{\rm{eff}}$.}
\end{center}
\end{figure*}

Following the previous section, we have shown that, at early-times, the scalar field will follow the matter domination attractor solution irrespective of the presence of a scalar potential $V(\phi)$. At late-times, its evolution should be dominated by $V(\phi)$, leading to a departure from the matter dominated attractor solution. Therefore, we now try fixing the background evolution to match that of a standard flat $w$CDM scenario, such that
\begin{equation}{\label{hubbledesigner}}
H^{2}(a) = \frac{H_{0}^{2} E(a)}{\phi} \equiv \frac{H_{0}^{2}}{\phi} \PR{\Omega_{\rm{m}}a^{-3} + E_{\rm{eff}}(a)},
\end{equation}
\noindent where $\Omega_{\rm{m}}$ is the present-day fractional matter energy density, and the dark energy component will be fixed as
\begin{equation}{\label{effde}}
E_{\rm{eff}}(a) = \PC{1-\Omega_{\rm{m}}}e^{3\int_{a}^{1} (1+w_{\rm{eff}}) \rm{d} \ln a}.
\end{equation}
We will be assuming that the effective dark energy equation of state $w_{\rm{eff}}$ is a constant such that $w_{\rm{eff}} \geq -1$. We should be clear, however, that this is not a limitation of this procedure: it can be easily extended to a varying $w_{\rm{eff}}$ by providing a $w_{\rm{eff}}$ as a function of the scale factor $a$. We merely choose to do so in hope of finding analytic expressions for some of the observables in terms of the fundamental parameters of the theory. Therefore, we can now numerically evolve the scalar field just by using Eq.~(\ref{fieldfrw}) without evolving the Hubble parameter using Eq.~(\ref{friedmann}). We are also effectively parameterizing Eq.~(\ref{rhoeff}) so that our dark energy component's energy density matches a $w$CDM type and are not worried with its exact numerical evolution.

We then take the approximation of considering the scalar field potential to be determined by,
\begin{equation}{\label{designpot}}
 V(\phi) = 3 H_{0}^{2} \PC{1-\Omega_{\rm{m}}}e^{3\int_{a}^{1} (1+w_{\rm{eff}}) \rm{d} \ln a},
\end{equation}
meaning that we are considering that the main contribution to the effective dark energy density comes from the scalar field potential, with the scalar field dynamics being sub-dominant. With this approximation we also don't expect to affect the matter domination attractor solution at early times since, as seen before, the potential contribution to $\Omega_{\phi}$ is not relevant in the matter dominated regime.

To generate our numerical results we have fixed the initial value of the scalar field $\phi(z_i)$ and $\phi^{\prime}(z_i)$ to match the matter dominated attractor solution value at a redshift of $z_{i} = 1000$. In Fig.~\ref{figure2} we plot the evolution of $\phi$ and $w_{\rm{eff}}$ for different values of $\Omega_{\rm{m}}$, $w_{\rm{eff}}$, and $\omega_{\rm{BD}}$ by numerically solving Eq.~(\ref{fieldfrw}) and fixing the evolution of $H$ with Eq.~(\ref{hubbledesigner}). 

We note that, similarly to what we observed in the constant potential case, the presence of the dark energy component leads to a departure of $\phi$ from the matter domination attractor solution at late times, leading to a scalar field value higher than $\phi_{0}$ at the present. And Fig.~\ref{figure2} makes it clear that this departure happens earlier in time and is more significant the earlier the dark energy component starts to dominate at late-times (which happens the bigger $w_{\rm{eff}}$ is or the smaller $\Omega_{\rm{m}}$ is). This means that, the higher $w_{\rm{eff}}$ is, the more relevant the scalar field dynamics becomes. Hence, our designer approach breaks down if $w_{\rm{eff}}$ is much higher than $-1$. 

Looking at Eq.~(\ref{rhoeff}), one might be concerned about the numerical evolution of the effective dark energy density which we parameterized by Eq.~(\ref{effde}); we would probably not recover a flat cosmology today due to the contribution of the scalar field dynamics to the overall critical density of the Universe. If we were to compute $\rho_{\phi}$ numerically with Eq.~(\ref{rhoeff}), one could adjust the weight of the potential $V(\phi)$ to compensate for the dynamics of the scalar field and recover $\Omega_{\phi} = 1-\Omega_{\rm{m}}$ today. Hence, in effect, $V(\phi) = 3 H_{0}^{2} \overline{\Omega}_{\phi} a^{-3(1+w_{\rm{eff}})}$, where $\overline{\Omega}_{\phi} \ne \PC{1 - \Omega_{\rm{m}}}$ could be found by performing a simple binary search, for example. We will provide an approximation for this factor using our analytical solutions for $\phi$ in Appendix \ref{appendix2}.

We can also study the evolution of $w_{\rm{eff}}$ in Fig.~\ref{figure2}. We see that we again have a sharp transition from the matter domination attractor regime $w_{\rm{eff}} = -0.4$ value at early times to the value we fix $w_{\rm{eff}}$ to at late-times. The scale factor at which this transition happens can be estimated from
\begin{equation}{\label{adiscgen}}
 a_{\rm{disc}} \approx \PC{\frac{\Omega_{\rm{m}}} {1-\Omega_{\rm{m}}} \frac{1}{1+\omega_{\rm{BD}}}}^{-\frac{1}{3w_{\rm{eff}}}},
\end{equation}
making it clear that, the larger $w_{\rm{eff}}$ is and the earlier our dark energy component becomes relevant, the earlier this transition happens. Also, even though we don't show that explicitly, we recover the GR plus $w$CDM limit when we take $\omega_{\rm{BD}} \rightarrow \infty$, and $w_{\rm{eff}}$ should then be equal to the value we fix it to be throughout the whole evolution, since the discontinuity now happens earlier or may even be completely avoided.  

\section{\label{analytical}Analytical solutions for $\phi$} 

With our designer approach in hand, we can now proceed to find analytical approximations to the scalar field evolution which, in turn, can be used to construct approximations to our observables. We first consider the $\Lambda$CDM-like case and then generalize to an arbitrary (but constant) effective equation of state $w_{\rm{eff}}$.

\subsection{\label{yes1}$w_{\rm{eff}}=-1$}
We start by expressing the scalar field equation of motion, given by Eq.~(\ref{fieldfrw}), in terms of $\ln \hspace{1 mm} a$. We then simplify it by simultaneously neglecting the $\phi^{\prime \prime}$ and $\phi^{\prime 2}$ terms, yielding
\begin{equation}{\label{fieldfrwapprox1}}
 \frac{\phi^{\prime}}{\phi}\PC{1 - \frac{1}{2}\frac{\Omega_{\rm{m}}a^{-3}}{1 - \Omega_{\rm{m}} + a^{-3}\Omega_{\rm{m}}}} = \frac{4\PC{1-\Omega_{\rm{m}}}+a^{-3}\Omega_{\rm{m}}}{d\PC{1 - \Omega_{\rm{m}}}+a^{-3}\Omega_{\rm{m}}},
\end{equation}
where $d = \PC{2 \omega_{\rm{BD}} + 3}$. The solution for the scalar field will be a fully analytical expression, given by
\begin{equation}{\label{phisolsimpleminus1}}
 \phi(a) = \phi(a_i) g(a_i)^{-1} g(a),
\end{equation}
where $\phi(a_i)$ is the scalar field value at a high redshift $z_i$ set by the matter dominated attractor solution, or can be fixed to be $\phi_{0}$ at $a_i = 1$. The function $g(a)$ is given by
\begin{equation}{\label{ga}}
 g(a) = a^{\frac{2}{d}}\PC{2a^{3}\PC{1-\Omega_{\rm{m}}}+\Omega_{\rm{m}}}^{\frac{2}{3d}}
\end{equation}

\begin{figure}
\begin{center}
\includegraphics[scale = 0.50]{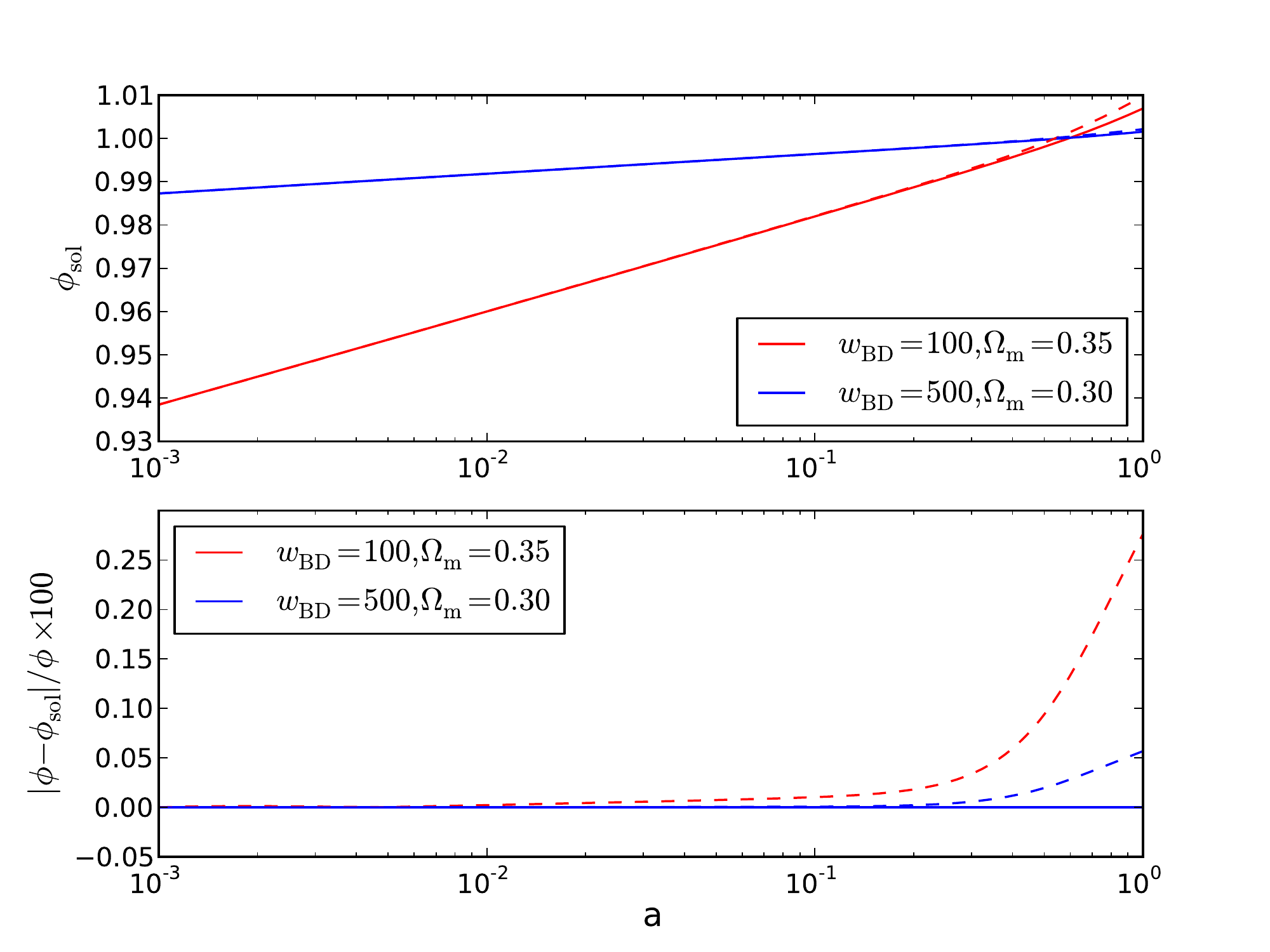}
\caption{\label{figure3}We show the analytical solution for the scalar field, $\phi_{\rm{sol}}$, predicted by Eq.~(\ref{phisolsimpleminus1}) (in solid lines) for different values of $\omega_{\rm{BD}}$ and $\Omega_{\rm{m}}$ in the upper plot. We compare the analytical solutions to the numerical evolution of $\phi$  (in dashed lines) predicted by our designer method, and show the relative error in the bottom plot. The errors are shown in \%.}
\end{center}
\end{figure}

We show the evolution of $\phi$ predicted by this solution in Fig.~\ref{figure3}. It exhibits a tendency to overestimate the deviation from the matter domination attractor solution at late-times. However, its errors are small, specially when considering the considerable simplification we have found to the full numerical analysis of our designer approach.

\subsection{\label{not1}$w_{\rm{eff}}\neq-1$}
We now extend our analytical approximation for cosmologies with $w_{\rm{eff}}\neq-1$. In these circumstances, we expect the dark energy component to become relevant earlier, and hence produce larger deviations from the matter dominated attractor prediction. We will focus mainly in the late-time evolution of $\phi$, when the dark energy component comes to dominate. For that effect, we re-express Eq.~(\ref{fieldfrw}) in terms of $\ln a$, and assuming $V_{\phi} = 1/\phi^{\prime}\hspace{0.5 mm}dV/d \ln a$, we approximate it as
\begin{equation}{\label{appeqmotion1}}
   \phi \frac{12 \phi^{\prime} \PC{1-\Omega_{\rm{m}}} + 18 \phi \PC{1-\Omega_{\rm{m}}}\PC{1+w_{\rm{eff}}}}{d \PC{1 - \Omega_{\rm{m}} + \Omega_{\rm{m}}a^{3w_{\rm{eff}}}}} \approx 3 \phi^{\prime 2},
\end{equation}
where $d = \PC{2 \omega_{\rm{BD}} + 3}$ and we have also neglected terms proportional to $\phi^{\prime \prime}$, $\phi^{\prime 3}$ and $(1+w_{\rm{eff}})(1-\Omega_{\rm{m}})$, the last two arising with the derivative of $H$. We have not included the matter driving term that dominates at early-times. Assuming that the driving term from the potential slope is much more significant than the $V(\phi)$ one- which effectively means $\phi^{\prime}$ is much smaller than unity for large $\omega_{\rm{BD}}$- we take the square root of this equation and perform a Taylor expansion of the left-hand side, obtaining:

 \begin{equation}{\label{appeqmotion2}}
 \frac{6\phi^{\prime}\PC{1-\Omega_{\rm{m}}}}{\sqrt{18\PC{1-\Omega_{\rm{m}}}\PC{1+w_{\rm{eff}}}}} + \sqrt{18\PC{1-\Omega_{\rm{m}}}\PC{1+w_{\rm{eff}}}}\phi - \phi^{\prime} \sqrt{3d\PC{1 - \Omega_{\rm{m}} + \Omega_{\rm{m}}a^{3w_{\rm{eff}}}}}\approx 0
 \end{equation}
 
\begin{figure}[t!]
 \begin{center}
\includegraphics[scale = 0.50]{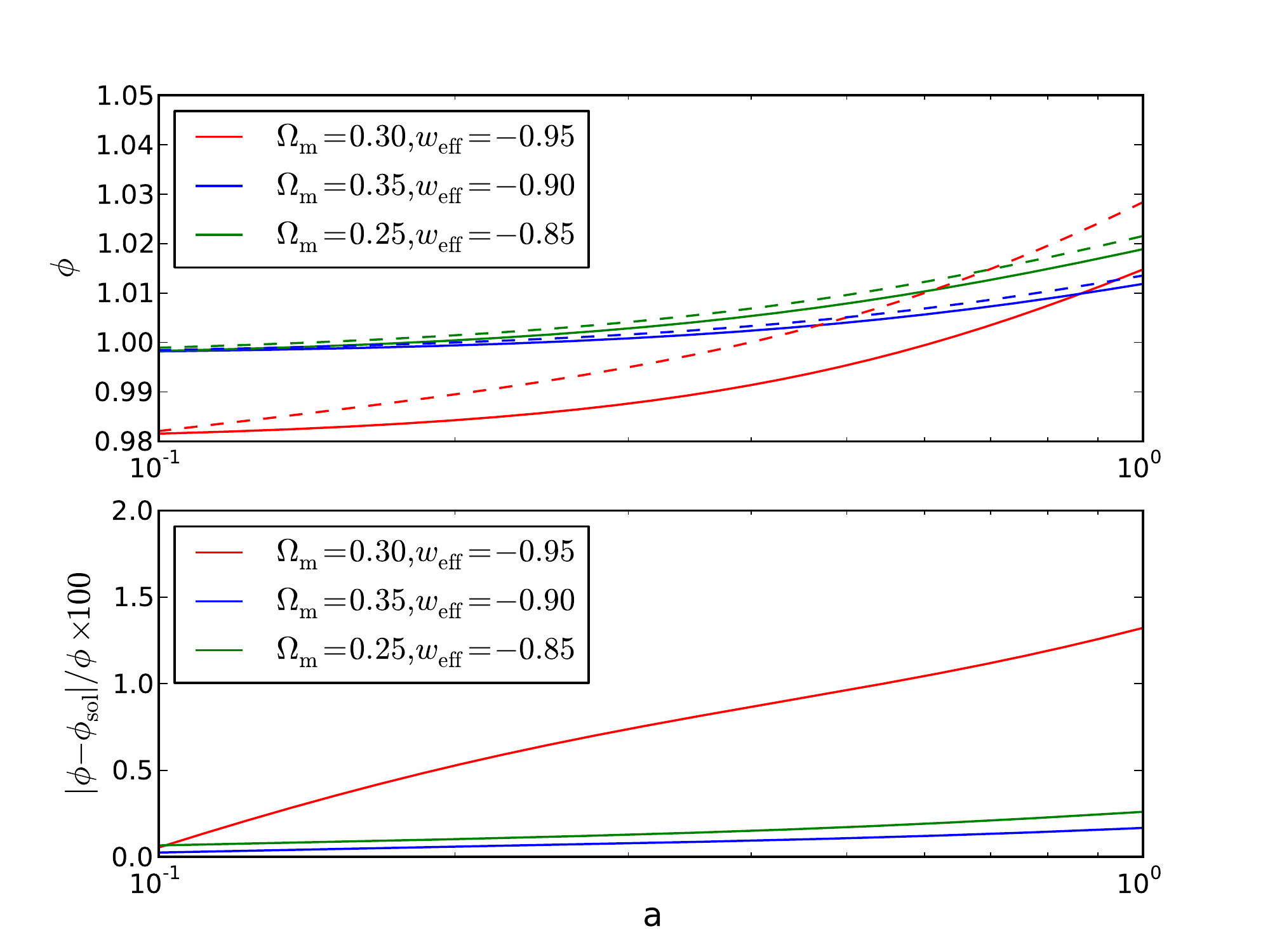}
\caption{\label{figuresol2}We show the late-time analytical solution (solid lines) of the scalar field predicted by Eq.~(\ref{phi_sol_intermediate}), $\phi_{\rm{sol}}$, when $w_{\rm{eff}} \neq -1$ for different values of $\omega_{\rm{BD}}$ and $\Omega_{\rm{m}}$. We compare it to the numerical solution (dashed lines) predicted by our designer approach in the top plot. We show the relative errors in the bottom plot, in \%. The red lines are for $\omega_{\rm{BD}} = 100$, while the green and blue lines are for $\omega_{\rm{BD}} = 1000$.}
\end{center}
\end{figure}

\noindent With these approximations, the solution for this equation is given by
\begin{equation}{\label{phi_sol_intermediate}}
 \phi(a) = \phi(a_i) f(a) f(a_i)^{-1},
\end{equation}
where $\phi(a_i)$ is the value of the scalar field at a desired scale factor $a_i$. This can either be set to the matter dominated attractor solution at a redshift $z_i \approx 10$ or to $\phi_0$ at $a = 1$ if one wants to fix the present-day value of the scalar field to recover $G_{\rm{eff}}/G = 1$ today. The function $f(a)$ is given approximately by
\begin{equation}{\label{functionofa}}
 f(a) \approx \PC{\frac{1+x}{x-1}}^{-\frac{\sqrt{6}\sqrt{d}\PC{1+w_{\rm{eff}}}^{3/2}}{w_{\rm{eff}}\PC{-2+3d(1+w_{\rm{eff}})}}},
\end{equation}
where $x = \sqrt{1 + \frac{\Omega_{\rm{m}}}{1-\Omega_{\rm{m}}}a^{3w_{\rm{eff}}}}$ and we have neglected similar terms whose exponents were proportional to $d^{-1}$. In Fig.~\ref{figuresol2} we compare the late-time evolution of $\phi$ predicted by Eq.~(\ref{phi_sol_intermediate}) with the numerical evolution found in our designer approach. We do so by fixing $\phi(a_i)$ to the matter domination attractor solution at $z_i = 10$ for all the cases. We see that this solution works better for larger $\omega_{\rm{BD}}$. Nevertheless, even if the agreement with the numerical solution is not perfect, the errors are small, and the overall form of $\phi$ is excellent for such a simple approximation.

\begin{figure}[t!]
\begin{center}
\includegraphics[scale = 0.50]{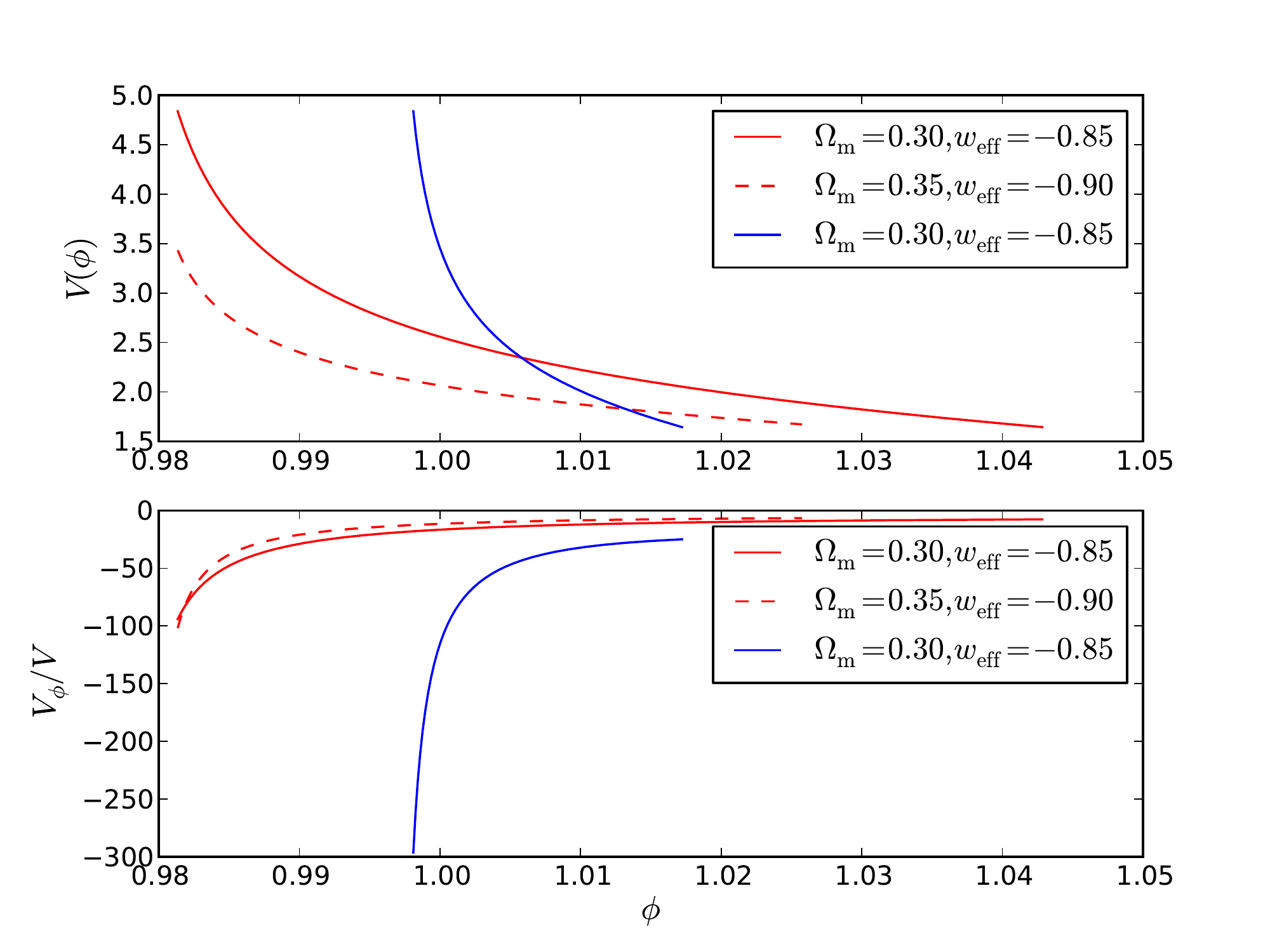}
\caption{\label{figpot}We show the reconstructed functional form of $V(\phi)$. The red lines have $\omega_{\rm{BD}} = 100$, while the blue line has $\omega_{\rm{BD}} = 1000$.}
\end{center}
\end{figure}

We are now also in a position to reconstruct the effective form of the self-interaction potential $V(\phi)$ across the entire cosmological evolution. For that, we invert the solutions for $\phi$ to get the scale factor as a function of the scalar field. We use the field's matter dominated attractor solution at early-times and our analytical approximation at late-times. Hence, the potential will be given by

\begin{equation}{\label{potentialphi}}
  V(\phi)=\begin{cases}
    3 H_0^{2} (1 - \Omega_{\rm{m}}) \PC{\phi/\phi_{0}}^{-3(1+w_{\rm{eff}})(1+\omega_{\rm{BD}})}, \hspace{3 mm} \text{during matter domination} & \\
    3 H_0^{2} (1 - \Omega_{\rm{m}}) \PR{ \frac{2^{2/3}\PC{1-\Omega_{\rm{m}}}^{1/3}\PC{\phi / c}^{\frac{\PC{-2+3d(1+w_{\rm{eff}})}w_{\rm{eff}}}{3\sqrt{6}\sqrt{d}\PC{1+w_{\rm{eff}}}^{3/2}}}}{\Omega_{\rm{m}}^{1/3}\PC{ 1 - \PC{\phi / c}^{\frac{\PC{-2+3d(1+w_{\rm{eff}})}w_{\rm{eff}}}{\sqrt{6}\sqrt{d}\PC{1+w_{\rm{eff}}}^{3/2}}}}^{2/3}}         }^{\frac{-3(1+w_{\rm{eff}})}{w_{\rm{eff}}}}, \hspace{3 mm} \text{at late-times},&
  \end{cases}
\end{equation}
where $c = \phi(a_i) f(a_i)^{-1}$, as defined in Eq.~(\ref{phi_sol_intermediate}). We plot the late-time form of the potential $V(\phi)$ and $V_{\phi}/V$ in Fig.~\ref{figpot} for different values of $\omega_{\rm{BD}}$, $w_{\rm{eff}}$ and $\Omega_{\rm{m}}$. We can observe that $V(\phi)$ exhibits a simple form, as in a standard run-away potential, with the slope decreasing at higher values of $\phi$ or, equivalently, close to the present. We see as well that $V_{\phi}/V$ takes significantly high, absolute values. This justifies our assumption in considering just the effect of the slope of the potential in the evolution of $\phi$ when $w_{\rm{eff}} \ne -1$. Indeed, this is the term that will have the most effect on the scalar field dynamics, leading to a significant departure from the attractor solution at late-times for $w_{\rm{eff}}>-1$.

In Fig.~\ref{figpot} we can also observe that the slope of the potential becomes more significant for higher values of $\omega_{\rm{BD}}$. This seem to contradict what we have seen in Fig.~\ref{figuresol2}, where the scalar field dynamics seem to be more relevant, the smaller $\omega_{\rm{BD}}$ is. However, the source terms for the evolution of $\phi$ are suppressed by a factor proportional to $\omega_{\rm{BD}}^{-1}$. Hence, for a larger value of $\omega_{\rm{BD}}$, the only way to have significant field dynamics at late-times, and hence induce a significant departure from the matter dominated attractor solution that produces a $w_{\rm{eff}} \ne -1$, is to have a very large source term. Finally, we can also see how, for larger $w_{\rm{eff}},$ we recover a more tilted potential: the more relevant we set our dark energy component to be, the more significant we expect the scalar field dynamics to be at late-times.

\subsection{\label{global}A global solution}

In the previous two sections, we presented solutions for the evolution of the scalar field that worked well for $w_{\rm{eff}} = -1$ and $w_{\rm{eff}}>-1$ separately. We will now propose an approximate global solution: \begin{equation}{\label{globalsol}}
 \phi_{\rm{global}}(a) = \phi(a_i) f(a)  g(a)  f(a_i)^{-1} g(a_i)^{-1},
\end{equation}
which is just the product of the solutions we previously found for $w_{\rm{eff}} = -1$ and for $w_{\rm{eff}} > -1$.

Note that when $w_{\rm{eff}} = -1$, we have that $f(a_i)=f(a)=1$. Hence, $\phi_{\rm{global}}(a)$ will be the exact solution for the scalar field equation of motion under the assumptions we discussed in Sec.~\ref{yes1}.
When $w_{\rm{eff}} > -1$, we note that the main contribution will come from the $f(a)$ and $f(a_i)$ terms; we already have seen in Section \ref{not1} how the scalar field dynamics are more significant when $w_{\rm{eff}} > -1$. Not only that, but we note that $d g(a) /d \ln a \propto d^{-1}$, whereas $d f(a) / d\ln a$ produces terms proportional to $\PC{\sqrt{d}}^{-1}$. Hence, assuming $d g(a)/ d \ln a << d f(a)/ d \ln a$ when $w_{\rm{eff}} > -1$, $\phi_{\rm{global}}(a)$ will be a solution of Eq.~(\ref{appeqmotion2}). We will use this full solution in the following sections for the phenomenological parameters, and show that it is indeed a good approximation for the overall behavior of the scalar field.

\section{\label{phenom}A model for the phenomenological parameters.}

It has been shown that, at the level of the background and linear cosmological perturbation theory, it is possible to completely characterize any modified theory of gravity in terms of a handful of time dependent functions \cite{tessapar}. We proceed to do so with our designer extended Brans-Dicke gravity. We have already discussed two of our time dependent functions: the time varying Newton's constant (at the level of the background), $G_0 = G/\phi$ and the effective equation of state, $w_{\rm{eff}}$. 

For linear perturbations, following the notation of Ref.~\cite{defelice}, we consider a perturbed metric about the FLRW background in the Newtonian gauge,
\begin{equation}{\label{pertmetric}}
 ds^{2} = -(1 + 2\Psi)dt^2 + a^{2}(t)(1 + 2\Phi)\delta_{ij}dx^i dx^j,
\end{equation}
where $\Psi$ and $\Phi$ are the scalar perturbations that we will refer to as Newtonian potentials and are decomposed as a series of Fourier modes of scale $k$ $(h/\rm{Mpc})$.

If we are interested in the the impact of matter perturbations on  galaxy and weak lensing surveys, we can focus on the modes that are well within the Hubble radius, i.e. such that the condition $k^2/a^2 \gg H^2$ is respected. In this {\it quasi-static} regime the evolution equation for the matter density perturbation $\delta_{\rm{m}}$ can be approximated as \cite{matter1}
\begin{equation}{\label{mattpert}}
 \delta_{\rm{m}}^{\prime \prime} + \PC{2 + \frac{H^{\prime}}{H}} \delta_{\rm{m}}^{\prime}- \frac{3}{2} \frac{G_{\rm{eff}}}{G} \Omega_{\rm{m}}(a) \delta_{\rm{m}} \simeq 0,
\end{equation}
where $\Omega_{\rm{m}}(a) = \rho_{\rm{m}}/3H^2$, and $G_{\rm{eff}}/G$ will be dependent on the model. The primes represent derivatives with respect to $\ln a$. In the extended Brans-Dicke theory, $G_{\rm{eff}}$ is given by \cite{defelice}
\begin{equation}{\label{geffbd}}
 \frac{G_{\rm{eff}}}{G} = \frac{1}{\phi}\frac{4 + 2\omega_{\rm{BD}} + 2\phi\PC{Ma/k}^2}{3 + 2\omega_{\rm{BD}} + 2\phi\PC{Ma/k}^2}, 
\end{equation}
where the $M$ term is \cite{defelice}
\begin{equation}{\label{msquared}}
 M^2 = V_{\phi \phi} + \frac{\omega_{\rm{BD}}}{\phi^3}\PR{\dot{\phi^2} - \phi \PC{\ddot{\phi} + 3H\dot{\phi}}}.
\end{equation}

At late times, when the dark energy component starts to become relevant, the mass term can be simplified and expressed in terms of the potential alone using the scalar field equation of motion, such that $M^2 \approx V_{\phi \phi} + \frac{V_{\phi}}{\phi}$ \cite{defelice}.
where $V_{\phi \phi}$ and $V_{\phi}$ correspond to the second and first order derivatives of the potential with respect to the scalar field, respectively.

One can also define the gravitational slip $\eta$ corresponding to the ratio between the two Newtonian potentials \cite{defelice}
\begin{equation}{\label{etabd}}
  -\frac{\Phi}{\Psi} \equiv \eta  = \frac{1 + \omega_{\rm{BD}} + \phi \PC{Ma/k}^2}{2 + \omega_{\rm{BD}} + \phi \PC{Ma/k}^2}
\end{equation}
\noindent which, again, should depend on the specifics of the scalar-tensor model. Lastly, the sub-horizon version of the Poisson equation can be written as \cite{defelice}
\begin{equation}
 \frac{k^2}{a^2} \Psi \simeq -4 \pi G_{\rm{eff}} \rho_{\rm{m}} \delta_{\rm{m}}.
\end{equation}

In standard GR, when we neglect matter shear, the anisotropy equation between the Newtonian potentials becomes a simple constraint equation, $\Psi = \Phi$, and $\eta$ should be $1$ throughout the cosmological evolution, as should $G_{\rm{eff}}/G$. Hence, in a modified gravity theory, a deviation in these parameters signals a departure from standard GR that can potentially be measured. From Eqs.~(\ref{geffbd}) and (\ref{etabd}) it is clear that the GR limit is recovered when $\omega_{\rm{BD}} \rightarrow \infty$, as expected, or when the field becomes supermassive and $M^2 \rightarrow \infty$. But we now wish to understand how these functions depend on time. To do so, it is convenient to study
\begin{eqnarray}
\label{subhorizonparams}
\xi_{QS} &\equiv&\lim_{k\rightarrow\infty} \frac{G_{\rm{eff}}}{G}= \frac{1}{\phi}\frac{4+2\omega_{\rm{BD}}}{3+2\omega_{\rm{BD}}}\nonumber \\
\eta_{QS}&\equiv&\lim_{k\rightarrow\infty} \eta=\frac{1+\omega_{\rm{BD}}}{2+\omega_{\rm{BD}}} 
\end{eqnarray}
and the inverse length scale
\begin{equation}{\label{massscale}}
k_M\equiv\sqrt{\frac{\phi}{1+\omega_{\rm{BD}}}}Ma.
\end{equation}

\begin{figure}[t!]
\begin{center}
\includegraphics[scale = 0.50]{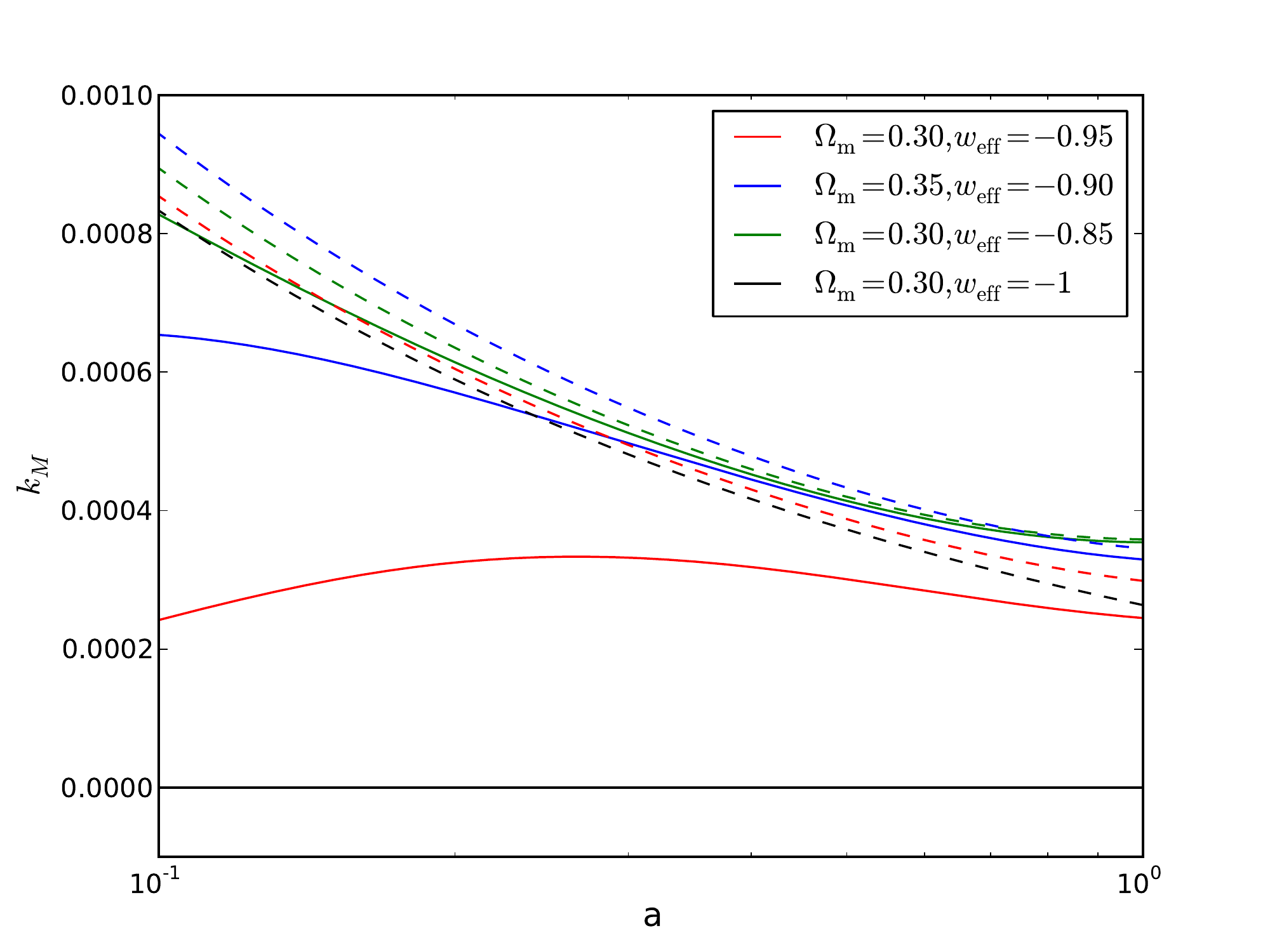}
\caption{\label{figure7}We show the evolution of $k_{M}$. The solid lines are the numerical evolution obtained using the global solution for $\phi$ given by Eq.~(\ref{globalsol}), while the dashed lines show the evolution predicted by the approximation given by Eq.~(\ref{massqs}). The red lines have $\omega_{\rm{BD}} = 1000$, the blue lines are for $\omega_{\rm{BD}} = 10000$ and the green lines have $\omega_{\rm{BD}} = 1 \times 10^5$. We have fixed $H_0 = 1/2997.9$ $h/\rm{Mpc}$, with $h=0.68$. The black lines are for $w_{\rm{eff}} = -1$, and the results shown are independent of $\omega_{\rm{BD}}$.}
\end{center}
\end{figure}
From Eq.~(\ref{subhorizonparams}), we see that $\eta_{\rm{QS}}$ is constant throughout the cosmological evolution, independently of the scalar field dynamics \cite{acquaviva,defelice,matter1}. Its GR limit is trivially recovered when we take $\omega_{\rm{BD}} \rightarrow \infty$. 

On the other hand, the late-time evolution of the mass scale parameter, $k_{M}$, can be written as:

\begin{equation}{\label{massqs}}
k_{M}(a) \approx \frac{3} {\sqrt{2}}H_{0}\sqrt{2\PC{1-\Omega_{\rm{m}}}\PC{1+w_{\rm{eff}}}a^{-\PC{1+3w_{\rm{eff}}}}+a^{-1}\Omega_{\rm{m}}\PC{2+w_{\rm{eff}}}}
\end{equation}
which is valid for $w_{\rm{eff}} > -1$ and large $\omega_{\rm{BD}}$. In the limit $w_{\rm{eff}} = -1$, this equation predicts a non-zero value for $k_{M}$, whereas it should be exactly zero throughout, as predicted by explicitly using our global solution for $\phi$ in Eq.~(\ref{massscale}). This should be evident as $V_{\phi} = V_{\phi \phi} = 0$ when $w_{\rm{eff}} = -1$. This approximation works fairly well for small redshifts and better for larger $\omega_{\rm{BD}}$, as can be seen in Fig.~\ref{figure7}. 

We can also observe that $k_M$ is a fairly negligible quantity, corresponding to scales which are of order or greater than the cosmological scale. To do so we compare $k_{M}$ to the comoving horizon, $aH = H_0 \sqrt{\Omega_{\rm{m}}a^{-1} + (1-\Omega_{\rm{m}})a^{-\PC{1+3w_{\rm{eff}}}}}$. It is not hard to see that $k_{M}/(aH) \lesssim 1$. Hence, the scale at which $k/k_{M}$ becomes relevant is approximately the same at which the perturbations $k$-modes become sub-horizon, which is at the basis of our assumptions. Therefore, taking $k/k_{M}>>1$ is an excellent approximation on quasi-static scales.

To understand the parameter dependence of $\xi_{QS}$ we perform a Taylor expansion around $a = 1$ using our approximate global solution for the scalar field. We further simplify our functions by considering the two regimes of interest we observe in our models: one where we will have $\phi = \phi_{0}$ today if we intend to recover $G_{\rm{eff}} = G$ at the present; and another one where we do not recover $\phi_0$ at the present, meaning that, essentially, we instead recover the matter domination attractor solution for $\phi$ at early times given by Eq.~(\ref{attphi}). For the first case, we obtain:
\begin{figure}[t!]
\begin{center}
\includegraphics[scale = 0.50]{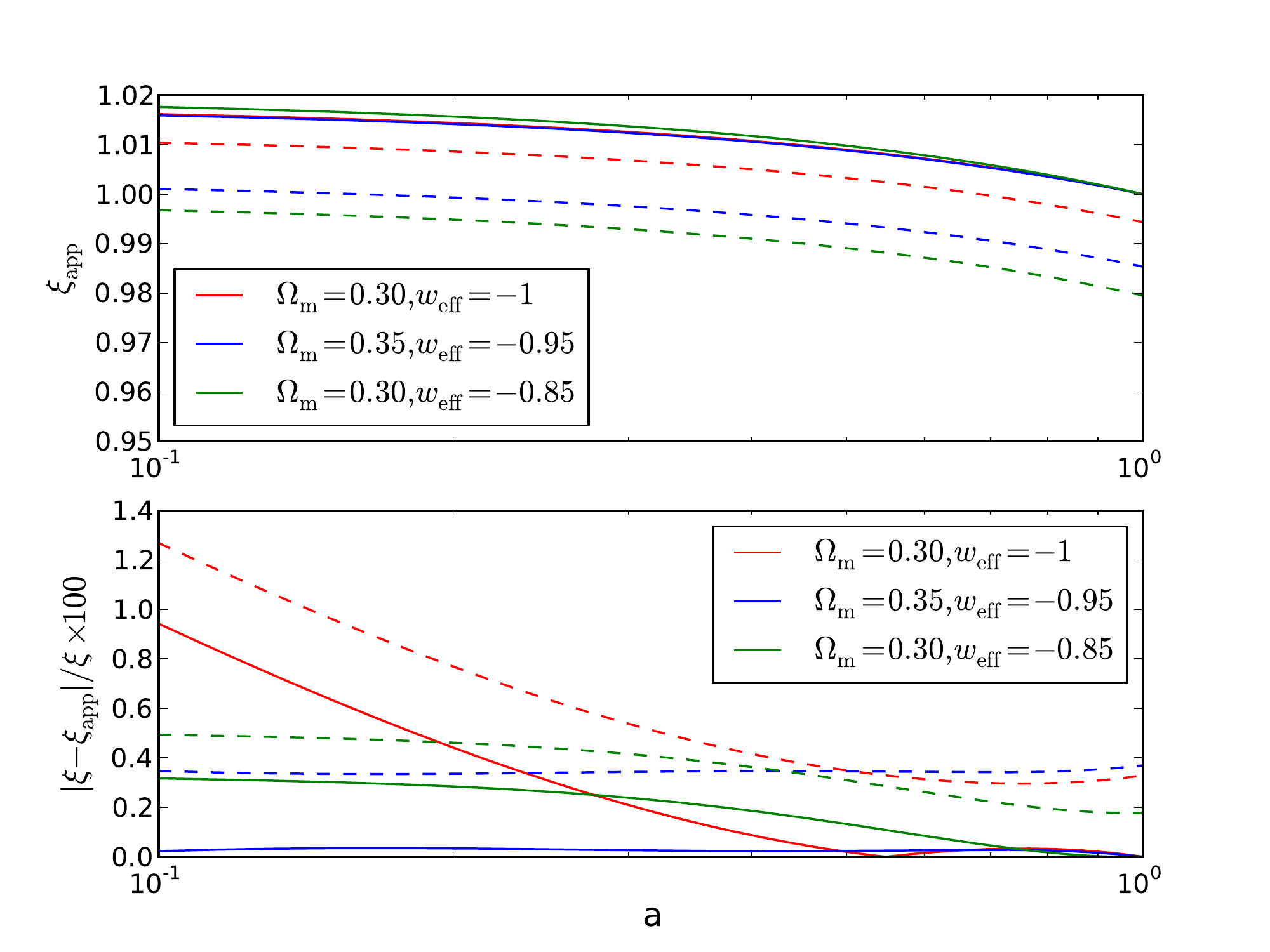}
\caption{\label{figure6}We show the late-time evolution of $G_{\rm{eff}}/G$ in the {\it quasi-static} regime using Eq.~(\ref{globalsol}) in the upper plot. The solid lines show $G_{\rm{eff}}/G$ with $\phi = \phi_{0}$ today, while the dashed lines do not have that restriction. The red lines are for $\omega_{\rm{BD}} = 100$, the blue lines for $\omega_{\rm{BD}} = 500$ and the green lines have $\omega_{\rm{BD}} = 1000$. The bottom plot shows the relative error of our approximations to the exact numerical solutions in \%.}
\end{center}
\end{figure}

\begin{equation}{\label{xiqsapprox1}}
\xi_{QS_1} (a) \approx 1 + \PC{1-a} \PR{\frac{8-6\hspace{0.1 mm}\Omega_{\rm{m}}}{d\PC{2-\Omega_{\rm{m}}}} + \frac{3\sqrt{6}\sqrt{d\PC{1-\Omega_{\rm{m}}}}\PC{1+w_{\rm{eff}}}^{3/2}}{3d\PC{1+w_{\rm{eff}}}-2}}
\end{equation}
while for the second case we find
\begin{equation}{\label{xiqsapprox2}}
 \xi_{QS_2} (a) \approx \PC{\frac{\Omega_{\rm{m}}}{2-\Omega_{\rm{m}}}}^{\frac{2}{3d}}\PC{\frac{\sqrt{\frac{1}{1-\Omega_{\rm{m}}}}-1}{1+\sqrt{\frac{1}{1-\Omega_{\rm{m}}}}}}^{-\frac{\sqrt{6}\sqrt{d}\PC{1+w_{\rm{eff}}}^{3/2}}{\PC{-2 + 3d\PC{1+w_{\rm{eff}}}}w_{\rm{eff}}}}\xi_{QS_1}(a)
\end{equation}
The GR limit of $\xi_{QS} = 1$ is recovered in both situations when we take $\omega_{\rm{BD}} \rightarrow \infty$. For the first case, as expected, $\xi_{QS} = 1$ when $a = 1$ since we have $\phi = \phi_0$ today. In the second case, $\xi_{QS} < 1$ today since the present-day value of the scalar field in these circumstances will always be larger than $\phi_0$. This can be observed in Fig.~\ref{figure6}, where we compare these approximations to the exact numerical solution of $\xi$ and we see they work considerably well.

We can now try and understand the dependence of $\xi_{QS}$ on the different parameters. Looking at $d = 2\omega_{\rm{BD}} + 3$, it becomes clear that increasing the Brans-Dicke parameter leads to $\xi_{QS}$ becoming closer to $1$ throughout the late-time cosmological evolution: its slope at $a = 1$, as given by Eq.~(\ref{xiqsapprox1}), decreases since it depends on the inverse of $d$ or $\sqrt{d}$. Then, looking at Eq.~(\ref{xiqsapprox2}), we see that the present-day value of $\xi_{QS}$ increases towards $1$ due to the exponents of the terms shown becoming extremely small.

Looking at the dependence of $\xi_{QS}$ on $\Omega_{\rm{m}}$, we realize it is similar to that on $\omega_{\rm{BD}}$. Increasing $\Omega_{\rm{m}}$ leads to both the slope of $\xi_{QS}$ decreasing in Eq.~(\ref{xiqsapprox1}) as well as the present day-value tending to $1$ in Eq.~(\ref{xiqsapprox2}). In Eq.~(\ref{xiqsapprox2}) we also see that, remarkably, our approximation recovers the matter dominated attractor solution value of $\xi_{QS} = 1$ when $\Omega_{\rm{m}} \rightarrow 1$.

Lastly, we have the effective equation of state parameter, $w_{\rm{eff}}$. Looking at Eq.~(\ref{xiqsapprox1}), we see that the slope of $\xi_{QS}$ will increase as $w_{\rm{eff}}$ becomes less negative, making its evolution more noticeable for larger $w_{\rm{eff}}$ when all other parameters remain fixed. Also, the exponent of the second term in Eq.~(\ref{xiqsapprox2}) increases for large $w_{\rm{eff}}$, leading to a significant departure of $\xi_{QS}$ today, producing values of $\xi_{QS} (a=1)$ that are detectably smaller than $1$ in a clear departure from standard GR. This is a reflection of the effect of increasing $w_{\rm{eff}}$ on the evolution of the scalar field $\phi$: the higher $w_{\rm{eff}}$ is, the sooner $\phi$ departs from the matter domination attractor solution and the larger its present-day value will be.

\begin{figure}[t!]
\begin{center}
 \includegraphics[scale=0.45]{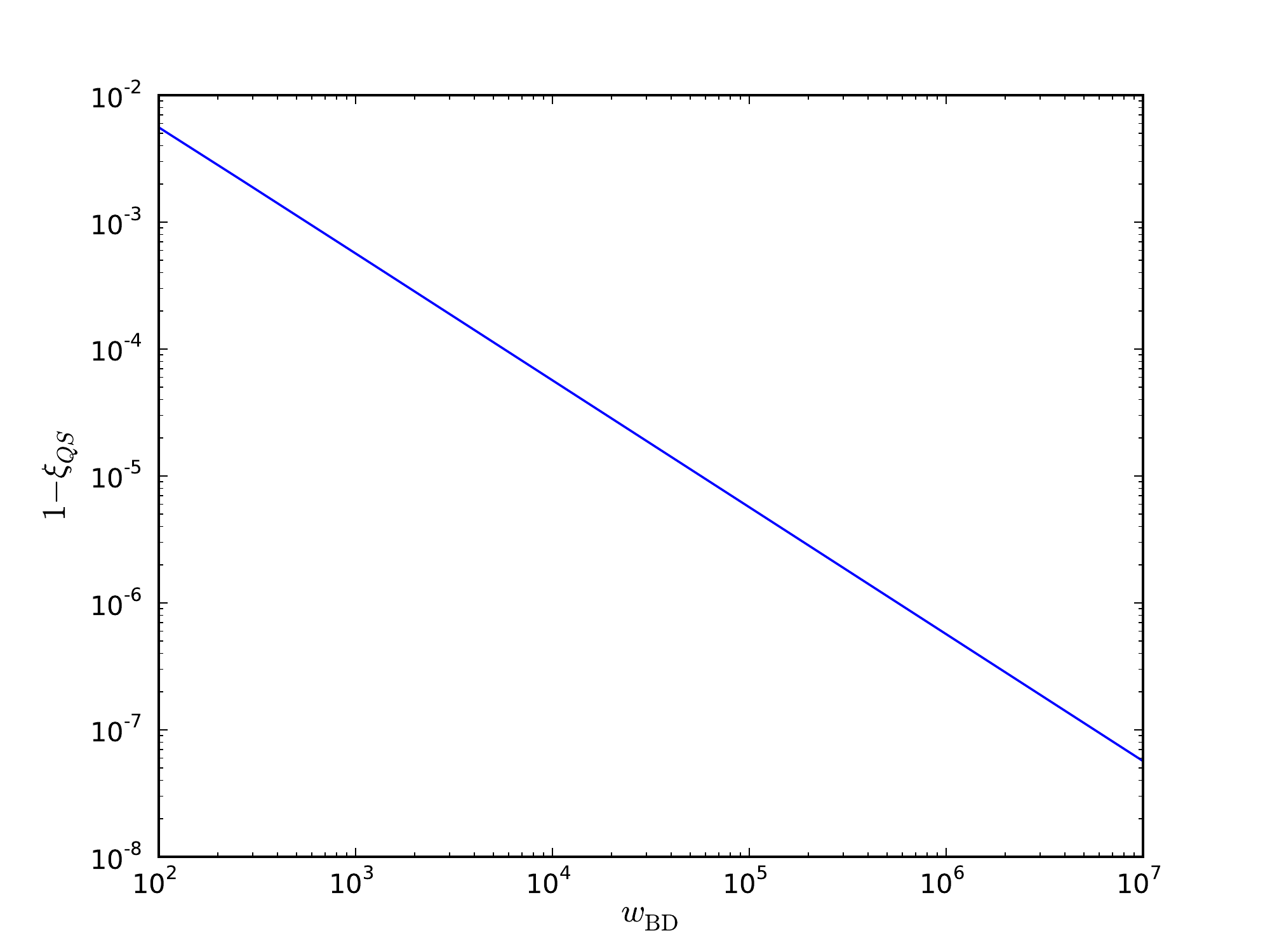}
 \caption{\label{figure8}We show the evolution of $\xi_{QS}$ at a = 1 for $\Omega_{\rm{m}} = 0.308$ and $w_{\rm{eff}} = -1$ as a function of $\omega_{\rm{BD}}$. For this plot we have used Eq.~(\ref{xiqsapprox2}), therefore assuming that $\phi(a=1)$ may not be equal to $\phi_0$.}
 \end{center}
\end{figure}

In Fig.~\ref{figure8}, we plot $\xi_{QS}$ as a function of $\omega_{\rm{BD}}$ at $a = 1$, using Eq.~(\ref{xiqsapprox2}). We see that if we don't fix $\phi = \phi_0$ today, there is a significant, possibly detectable, deviation from the standard GR  value, $\xi_{QS} = 1$, even for very large $\omega_{\rm{BD}}$. Of course, we also see that when $\omega_{\rm{BD}} \rightarrow \infty$, $\xi_{QS}$ tends to $1$. Therefore, in order to be competitive with Solar System constraints $\omega_{\rm{BD}} > 10^{4}$ \cite{will,bertotti}, we would have to able to measure $\xi_{QS}$ with a precision of around $10^{-4}$.

\section{\label{discussion}Discussion}

In this paper we have applied the designer approach to the extended Brans-Dicke theory with the explicit presence of a self-interacting potential $V(\phi)$. By fixing the expansion history to that of an effective $w$CDM dark energy model, we are able to retrieve the scalar field evolution under the assumption that the main contribution to the effective dark energy density comes from the potential $V(\phi)$.

The numerical solutions we obtain have the property of respecting the matter domination attractor solution of Brans-Dicke models at early-times. At late-times, the scalar field departs from this solution and evolves more rapidly and towards larger values, yielding a value today larger than $\phi_0$, where $\phi_0$ is the present-day value of the matter regime attractor solution that ensures that one would measure $G_{\rm{eff}}$ today equal to the actual Newton's gravitational constant, $G$, in a matter-dominated Universe. This transition from the attractor solution happens earlier whenever we take a larger dark energy equation of state, $w_{\rm{eff}}$. However, if we constrain the present-day value of $\phi$ to be equal to $\phi_0$, our numerical solutions follow the power-law behavior of the attractor solution, shifted towards smaller values at early-times. When the evolution departs from the matter dominated behavior, we are then able to recover $\phi(a=1) = \phi_0$ as intended. 

We were able to obtain separate analytical approximations for the evolution of the scalar field when $w_{\rm{eff}} = -1$ and $w_{\rm{eff}}>-1$, which we then used to construct a global solution valid for $w_{\rm{eff}} \geq -1$. These approximations work remarkably well, with errors of sub-percent  for large values of $\omega_{\rm{BD}}$. These approximations also allowed us to reconstruct the late-time functional form of the potential $V(\phi)$; we found a simple run-away potential whose slope is inevitably dependent on $w_{\rm{eff}}$ and $\omega_{\rm{BD}}$. We reiterate that we have limited our analysis to constant $w_{\rm{eff}}$ so as to obtain analytical solutions which will shed light on the parameter dependence of the various observables we are considering; a non-constant  $w_{\rm{eff}}$ will severely complicate any attempts at doing so. However, we stress that the numerical implementation of the designer approach presented in Sec. \ref{designerbd} can be easily extended to a non-constant $w_
{\rm{eff}}$.

With these analytic approximations in hand, we then focused on the phenomenological parameters that describe the sub-horizon evolution of the linear perturbations of the theory.  We showed how the effective scale of the theory, which we designated by $k_{M}$, is of order the cosmological horizon; as a result we find that there is negligible scale dependence of the phenomenological parameters on observable scales. We found that the ratio between the Newtonian potentials, $\eta = \Psi/\Phi$ is constant throughout the cosmological evolution, for large values of the Brans-Dicke parameter \cite{defelice}. We also found simple analytical expression for $\xi = G_{\rm{eff}}/G$ which depend explicitly on the parameters of the theory, as seen in Eqs.~(\ref{xiqsapprox1}) and (\ref{xiqsapprox2}).

One of the main features of this model is the possibility of having $\xi_{QS} \ne 1$ today; this is due to the departure of the scalar field from the matter dominated attractor solution at late-times such that its present-day value will be larger than $\phi_0$. The present-day value of $\xi_{QS}$ at $a = 1$, given in Eq.~(\ref{xiqsapprox2}) tends to $1$ as $\omega_{\rm{BD}} \rightarrow \infty$ since the exponent of the terms shown tend to zero. Also, as for $w_{\rm{eff}}>-1$, the exponent of one of the terms increases, leading to smaller values of $\xi_{QS}$ today, even when $\omega_{\rm{BD}}$ is very large. 

If, however, we impose $\xi_{QS}$ to be $1$ today, its evolution is predicted by Eq.~(\ref{xiqsapprox1}). In these circumstances, the main distinguishing point between this model and standard GR will be the slope of $\xi_{QS}$ at the present: for the extended Brans-Dicke theory it can be different from zero. We note that, even when $w_{\rm{eff}} = -1$, the predicted slope is different from zero. Hence, even the simple extended Brans-Dicke+$\Lambda$CDM model could be ruled out if $\xi_{QS}$ is found to not vary close to the present.

Finally, we note that in order to attain constraints on $\omega_{\rm{BD}}$ that are competitive with those obtained in Solar-system tests \cite{will,bertotti}, $\xi_{QS}$  and $\eta_{QS}$ would naively need to be constrained with a precision of around $10^{-4}$. This is a formidable challenge, but one should bear in mind that $\eta_{QS}\neq1$ throughout (at least) the matter dominated era while the same is possible for $\xi_{QS}$. This means that there will be a cumulative effective (as shown in \cite{Baker,Leonard}) which means that constraints on the growth rate (or weak lensing) of order $10^{-3}$ or even $10^{-2}$ might be sufficient to place competitive constraints on $w_{\rm BD}$. 

\acknowledgments

We would like to thank Andrew Liddle and Tessa Baker for helpful discussions and comments on this paper. N.A.L.\ also acknowledges financial support from Funda\c{c}\~{a}o para a Ci\^{e}ncia e a Tecnologia (FCT) through grant SFRH/BD/85164/2012. P.G.F.\ acknowledges support from STFC, BIPAC, a Higgs visiting fellowship and the Oxford Martin School.

\small
\appendix

\section{\label{appendix2}Correction factor for V($\phi$)}
In this appendix we show the correction factor one can add to the potential $V(\phi)$ in order to balance the scalar field dynamics in the exact numerical evolution of $\rho_{\phi}$ in order to recover a flat Universe today. Effectively, we want to solve the equation
\begin{equation}
 -\frac{\phi^{\prime}(a_0)}{\phi(a_0)} + \frac{\omega_{\rm{BD}}}{6} \PC{\frac{\phi^{\prime}(a_0)}{\phi(a_0)}}^2 + \frac{1 - D\Omega_{\rm{m}}}{\phi(a_0)} = (1 - \Omega_{\rm{m}}),
\end{equation}

\noindent where $a_0 = 1$ and $D$ will be the correction factor such that $1 - D \Omega_{\rm{m}} \equiv \overline{\Omega}_{\phi}$, as discussed in Sec. \ref{designerbd}. First we show that factor using our solution for $w_{\rm{eff}} = -1$ using Eq.~(\ref{phisolsimpleminus1}):

\begin{equation}{\label{corr1}}
 D = \frac{1}{\Omega_{\rm{m}}} + \frac{\phi(a_i)g(a_0)}{g(a_i)}\frac{\PR{3d\PC{\Omega_{\rm{m}}-2}\PC{8 - 6\Omega_{\rm{m}} + d\PC{2-\Omega_{\rm{m}}}\PC{1-\Omega_{\rm{m}}}}+2\omega_{\rm{BD}}\PC{4-3\Omega_{\rm{m}}}^2}}{3\Omega_{\rm{m}}d^2\PC{\Omega_{\rm{m}}-2}^2},
\end{equation}
where $g(a)$ is defined in Eq.~(\ref{ga}) and $\phi(a_i)$ is the value of the scalar field at the starting redshift $a_i$.

Lastly, we show the correction factor for the case $w_{\rm{eff}} > -1$. For this part we have used the late-time solution for $\phi$ given by Eq.~(\ref{phi_sol_intermediate}):

\begin{eqnarray}
{\label{corr2}}
 D = \frac{1}{\Omega_{\rm{m}}} - \frac{\phi(a_i)}{\Omega_{\rm{m}}} c \Bigg( 1 - \Omega_{\rm{m}} + \frac{3\sqrt{6}\sqrt{d}\sqrt{1-\Omega_{\rm{m}}}\PC{1+w_{\rm{eff}}}^{3/2}}{-2 + 3d\PC{1+w_{\rm{eff}}}} - \frac{9\omega_{\rm{BD}}\PC{1-\Omega_{\rm{m}}}d\PC{1+w_{\rm{eff}}}^{3}}{\PC{2-3d\PC{1+w_{\rm{eff}}}}^{2}} \Bigg),
\end{eqnarray}
where $c =  f(a_0)/f(a_i)$, and $f(a)$ is defined by Eq.~(\ref{functionofa}).

\end{document}